\documentstyle[aps,prd,multicol,floats,epsfig,epsf]{revtex}
\begin{document} 
\draft
\wideabs{
\title{Thermodynamic control and dynamical regimes in protein folding}
\author {P.F.N. Faisca$^{1,2}$ and R.C. Ball$^{1}$ }
\address{$^{1}$Department of Physics, University of Warwick, Coventry
CV4 7AL, U.K.}
\address{$^{2}$CFMC, Av. Prof. Gama Pinto, 2, 1649-003 Lisboa Codex, Portugal}
\maketitle
\begin{abstract}

Monte Carlo simulations of a simple lattice model of protein folding
show two distinct regimes depending on the chain length. The first regime well
describes the folding of small protein sequences and its kinetic counterpart appears
to be single exponential in nature, while the second regime is typical of sequences
longer than 80 amino acids and the folding performance achievable is sensitive to
target conformation. The extent to which stability, as measured by the energy of a
sequence in the target, is an essential requirement and affects the folding
dynamics of protein molecules in the first regime is investigated.
The folding dynamics of sequences whose design stage was restricted to a certain
fraction of randomly selected amino acids shows that while some degree of stability is a
necessary and sufficient condition for successful folding, designing sequences that provide
the lowest energy in the target seems to be a superfluous constraint. 
By studying the dynamics of under annealed but otherwise freely designed sequences
we explore the relation between stability and kinetic accessibility.
We find that there is no one-to-one
correspondence between having low energy and folding quickly to
the target, as only a small fraction of the most stable sequences were
also found to fold relatively quickly.
\end{abstract}
}
\narrowtext

\section{Introduction}
In the early sixties, the Nobel Laureate Christian Anfinsen showed through in
vitro experiments that denaturated proteins can refold to their original native
structure in the absence of any catalyst {\cite {ANF}}. This suggested that
protein folding (PF) can be a spontaneous, first-order process, and that the only
information required for the protein sequence to fold correctly
is the sequence itself. Thus, in Anfinsen`s perspective,
sequence is the only determinant of the rates and mechanisms of folding.
Soon after these discoveries, Levinthal pointed out that a random search of the
conformational space, as implied by Anfinsen's thermodynamic hypothesis,
could not explain the time scale of folding as observed in Nature {\cite {LEV}}.
To bypass this paradox, Levinthal proposed a kinetic view, that proteins must
fold through some directed process, whose nature could involve for instance,
the existence of folding pathways. The latter do not necessarily imply a fixed
sequence of events in folding, nor do they require the existence of observable
folding intermediates.\par
\begin{figure}
\psfig{file=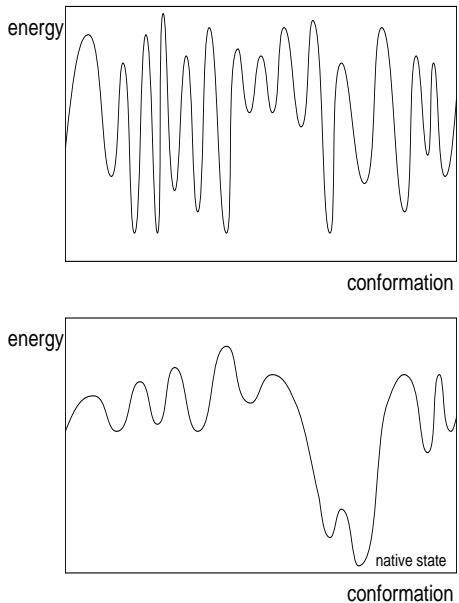, height=6cm, width=8cm, angle=270}
\caption{Energy landscape of a random heteropolymer (upper figure) and a
 protein landscape exhibiting a folding funnel.} 
\label{figure:no1}
\end{figure}

 
In the late 1980's, Brygenlson and Wolynes {\cite {BW}} introduced the concept
of energy landscape---the free energy as a function of protein conformation---as an
attempt to reconcil the kinetic and thermodynamic views of PF.
The 'topology' of the landscape characterizes the folding kinetics through the
existence of folding pathways. In Fig.~\ref{figure:no1} we show a cross section of
the energy landscape of a hypothetical random heteropolymer: as a consequence of the
existence of numerous local energy minima these sequences tend to behave as highly
frustrated systems. 
Nevertheless, it has been shown that a very small fraction of random sequences are able
to stably fold to what can be considered their native states (a deep energy minimum of the
energy landscape), in a biologically acceptable timescale {\cite {S0}}.\par
Dill and co-workers {\cite {D0}} proposed that a stable,
fast folding protein sequence must satisfy two essential requirements: 
thermodynamic stability meaning the existence of a deep global minimum in the
energy landscape and kinetic accessibility meaning the existence of a basin
of attraction sloping toward that minimum. 
This basin of attraction, first proposed by Leopold {\it {et al}} {\cite {LEO}} 
has become one of the most important concepts in protein folding dynamics and is
commonly known as the folding funnel.\par
In 1993, Shakhnovich and Gutin {\cite {SG}} developed a design method with the
purpose of creating protein like sequences, that is, sequences that fold fast and
stably to their respective native structures. The method is based on the
thermodynamic stability requirement, and was inspired by the behavior 
displayed by that very small fraction of protein like random sequences.
Shakhnovich {\cite {S}} claims that a thermodynamic driven
sequence selection actually solves the kinetic problem as well,
which in turn suggests a correlation between thermodynamic stability
and kinetic accessibility. This design method has been widely used to
study the folding dynamics of protein sequences whose length ranges from 
a few to at least 80 beads in simple lattice models. \par
Is it possible to fold longer protein chains ($>80$ amino acids) using the
thermodynamic stability as the unique driver of sequence design? This open question
was the starting point and initial motivation for the present paper.
As in previous studies, Monte Carlo simulations of a simple lattice model
were used to tackle the problem. In the present case we particularly focus on
the effects of native state structure on the folding dynamics.\par
This paper is organized as follows. Sec. II reviews the model as well
as the numerical methods used to sample both conformational and 
sequence space.
In Sec. III the numerical results are presented. We start by studying
the dependence of the folding time on temperature and then explore the 
extent to which the former is a sequence specific kinetic property. Then, 
we present evidence for the existence of two distinct dynamical regimes in PF.
Finally, for sequences which to fold in accordance to what we call the
first regime, we explore the effects of stability on folding dynamics and
also how it correlates with kinetic accessibility. In Sec. IV we make 
some final comments and conclusions.        

\section{Models and Methods \label{sec:2}}
\subsection{Folding in conformational space}
\begin{figure}
\psfig{file=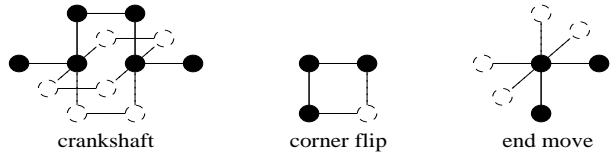, height=2cm, width=8cm, angle=360}
\caption{Move set used to generate the dynamics.}
\label{figure:no2}
\end{figure}

Part of the contribution of computational physics to protein folding
includes several findings obtained in the scope of simple lattice models.
Most generally, these models consider a coarse-grained description of the
protein, reduced to its backbone structure through a 'bead \& stick' 
representation.
Each bead is ascribed  a certain chemical identity representing an amino acid
type, while each stick stands for the peptide bond that covalently connects 
amino acids along the polypeptide chain. This structure is allowed to move
in a three dimensional infinite lattice, subjected to excluded volume 
constraints and exploring the conformational space in accordance with 
a kink-jump dynamics obeying the restriction 
that no bond length changes (Fig.~\ref{figure:no2}).
\begin{figure}
\psfig{file=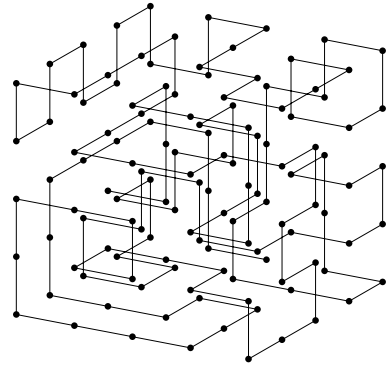, height=10cm, angle=90}
\caption{Example of a $125$ bead long target found by homopolymer relaxation.} 
\label{figure:no3}
\end{figure}
The energy of a conformation is given by the
contact Hamiltonian
\begin{equation}
H(\lbrace \sigma_{i} \rbrace,\lbrace \vec{r_{i}} \rbrace)=\sum_{i>j}^N
\epsilon(\sigma_{i},\sigma_{j})\Delta(\vec{r_{i}}-\vec{r_{j}}),
\label{eq:no1}
\end{equation}
\noindent
where $\lbrace \sigma_{i} \rbrace$ represents an amino acid sequence,
$\sigma_{i}$ standing for the chemical identity of bead $i$, while $\lbrace
\vec{r_{i}} \rbrace$ is the set of bead coordinates that define the conformation in
question. The contact function $\Delta$ equals $1$ if beads $i$ and $j$ are in
contact but not covalently linked and is $0$ otherwise. We follow many previous 
studies in taking the interaction parameters $\epsilon$ from the $20 \times 20$
Myazawa-Jerningan matrix, derived from the distribution of contacts in native
proteins {\cite {MJ}}.
Eq.~(\ref{eq:no1}) defines an energy function in conformational space
whose graph is the energy landscape. The energy landscape is explored via
the Metropolis Monte Carlo (MC) acceptance rule. This means that downhill
transitions (that lower the energy) are accepted with probability unity
and uphill transitions with probability proportional to the Boltzmann factor. 
Thus if $\Gamma_{A}$ and $\Gamma_{B}$ are two distinct conformations, and $\Delta
H=H_{A}-H_{B}$ is the energy difference between the states defined by these
conformations,
\begin{equation}
P_{ \Gamma_{A} \rightarrow \Gamma_{B}}= \left \{
\begin{array}{ll}
1 & \mbox{if }\Delta H < 0 \\
\exp \lbrack -\Delta H/k_B T \rbrack & \mbox{if }\Delta {H} \geq 0, 
\end{array}
\right.
\label{eq:no2}
\end{equation}
where $P$ is the transition probability, $T$  the temperature and $k_{B}$ the
Boltzmann constant.

\subsection{Design in sequence space}
The first step at the design stage is the choice of the target structure.
A target is a maximally compact and otherwise arbitrary structure such as the one
shown in Fig.~\ref{figure:no3}. Maximally compact structures maximize the 
number of contacts for a certain chain length and therefore minimize the degeneracy
of the Hamiltonian ~(\ref{eq:no1}).
The goal of the design process is, given a target,
to find a sequence that folds to it efficiently, and we follow Shakhnovich and
Gutin {\cite {SG}} in attempting this by seeking the sequence with the lowest possible 
energy in the target state, as given by Eq.~(\ref{eq:no1}). For this purpose, 
target's co-ordinates are quenched and the energy of Eq.~(\ref{eq:no1}) must be 
annealed with respect to the sequence variables. This naturally leads to the idea of simulated
annealing in sequence space; starting from a random initial sequence,
transitions between different sequences are successively attempted---different
sequences are generated by randomly permutating pairs of beads---along with a suitable
annealing schedule. If $S_{A}$ and $S_{B}$ are two different sequences the 
transition probability, $P_{ S_{A}\rightarrow S_{B}}$, comes given Eq.~(\ref{eq:no2})
with the temperature replaced by $T_{i}=\alpha T_{i-1}$ and $0< \alpha <1$.
This optimization procedure has been coined simulated annealing in
sequence space. It was the first successful procedure to design protein
sequences.

\section{{\it In virtuo} results \label{sec:3}}

\subsection{Finding the optimal folding temperature}
\begin{figure}[!h]
\psfig{file=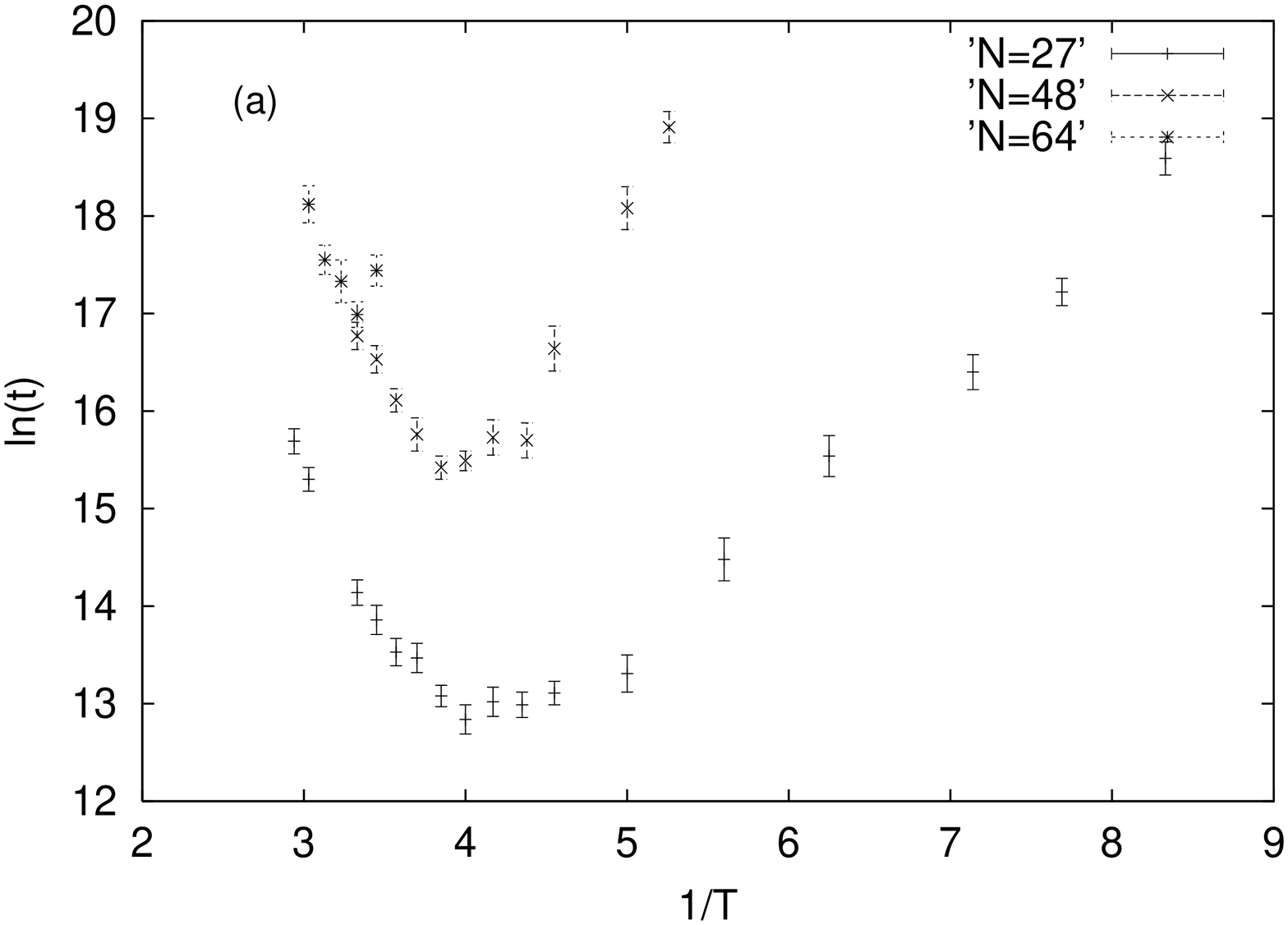, height=8cm, width=8cm, angle=270}
\psfig{file=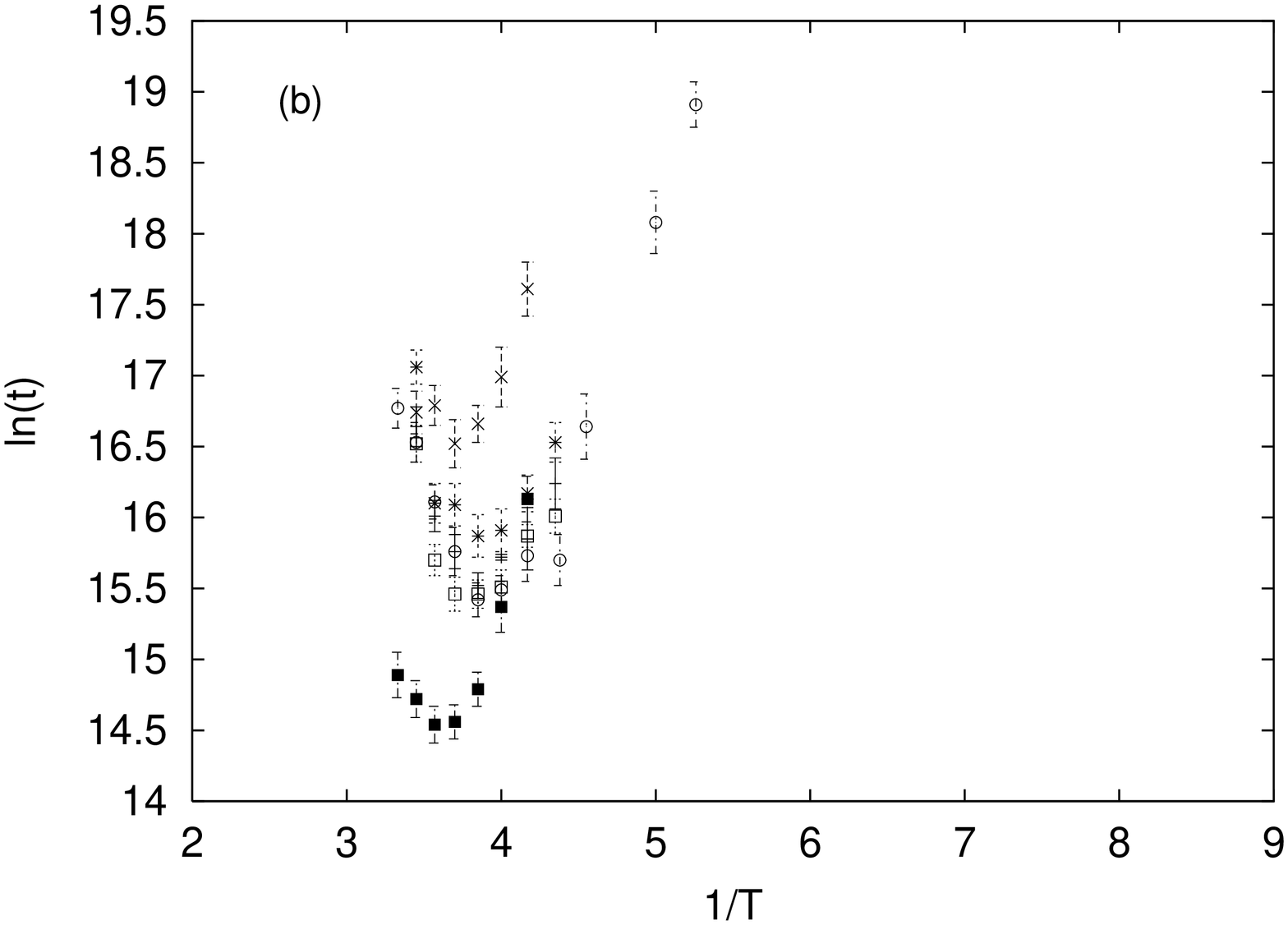, height=8cm, width=8cm, angle=270}
\caption{(a) Dependence of the folding time $t$ on the inverse temperature $1/T$ for
several values of the chain length.
Note that as $N$ increases the temperature range where folding can be observed
narrows. Fig.~\ref{figure:no4}(b) shows the same dependence for six
different 48 bead long sequences trainned to the same target.} 
\label{figure:no4}
\end{figure}

For each number of monomers $N=27$,$36$,$48$,$64$,$80$ and $100$
we found $5$ different maximally compact target structures by homopolymer relaxation. This method
is an efficient way to systematically find kinetically accessible maximally compact
structures and was previously used by Abkevich {\it {et al}} in {\cite{AB}} for
36 bead long targets.
Then, for each target we prepared a set composed of $30$ sequences
by applying the Shakhnovich method as previously outlined.
Next, we went on to determine the optimal folding temperature, $T_{fold}$.
For this purpose, a designed sequence at each $N$ was randomly selected and
subjected to MC folding simulations at several temperatures.
For $N=27,36$,$48$ and $64$ a set of 50 MC runs was performed
for each temperature and the folding time $t$ was taken as the value of
the mean first passage time (FPT) to the target averaged over the 50 MC runs.
Results plotted in  Fig.~\ref{figure:no4}(a) show that there is an optimal
folding temperature, $T_{fold}$, where folding to the target structure 
proceeds relatively fast. However, away from this optimum, both at higher 
and lower temperatures the process gets increasingly slower. In the first case
the protein sequence tends to behave like a random heteropolymer
rapidly fluctuating between unfolded states. In the second case the folding
kinetics gets slower because there is a high probability for the chain to get
trapped into metastable states and folding entails overcomming the corresponding
energy barriers {\cite {S1}}.\par
For $N=80$ and $N=100$ the number
of successful folding runs per each studied temperature, was only one half of the
attempted total. For this reason, we choose $T_{fold}$ as the temperature where the
highest ratio of folding success could be observed.\par
It has been claimed that the folding time and temperature are both sequence
specific parameters {\cite {SOCCI}}. To investigate this issue, five 48 bead long sequences
were randomly selected (one sequence per target) and their folding behavior was
studied over a temperature range as shown in Fig.~\ref{figure:no4}(b).
These show that whilst folding to the target can be quite target and (or) sequence dependent,
the optimal folding temperature is close to a self-averaging quantity in our simulations. It
should be emphasised that our sequence design preserved overall chemical composition, in 
contrast to earlier work {\cite {SOCCI}} which for unrestricted binary alphabet sequences found 
the optimal temperature to be more sequence dependent.

\subsection{Dependence of the folding probability on folding time: evidence for
two folding regimes.}
\begin{figure}
\psfig{file=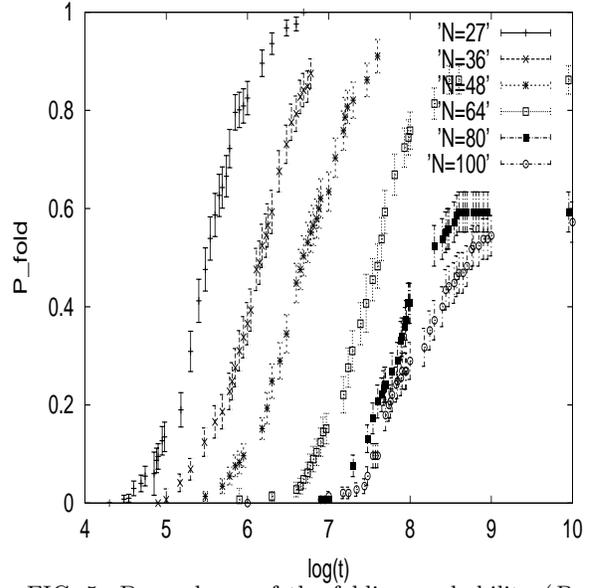, height=8cm, width=8cm, angle=270}
\caption{Dependence of the folding probability $(P_{fold})$ on $\log(t)$. For each
curve 150 simulations were used distributed across the available targets and sequences.
$P_{fold}$ was calculated as the number of folding simulations which ended up
to time $t$ normalized to the total number of runs.} 
\label{figure:no5}
\end{figure}

\begin{figure}
\psfig{file=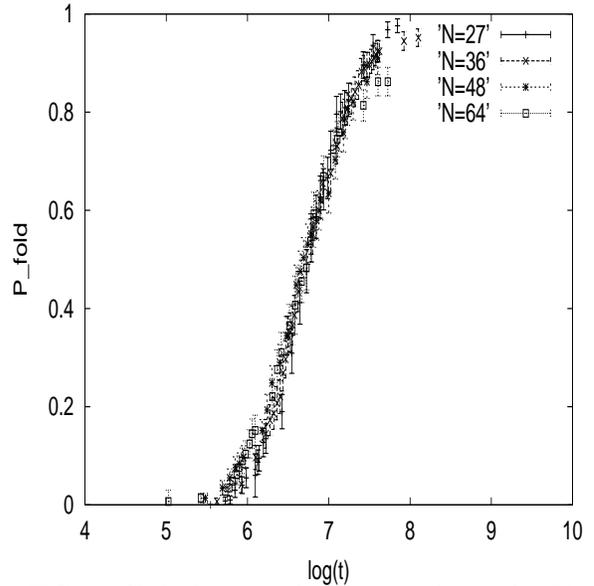, height=8cm, width=8cm, angle=270}
\caption{Shifted curves for $P_{fold}$ {\it vs} $\log(t)$ for $N=27$, $N=36$
and $N=64$. These curves were obtained by shifting the folding times
by a factor of $5\log(\frac{48}{N})$.}
\label{figure:no6}
\end{figure}

\begin{figure}
\psfig{file=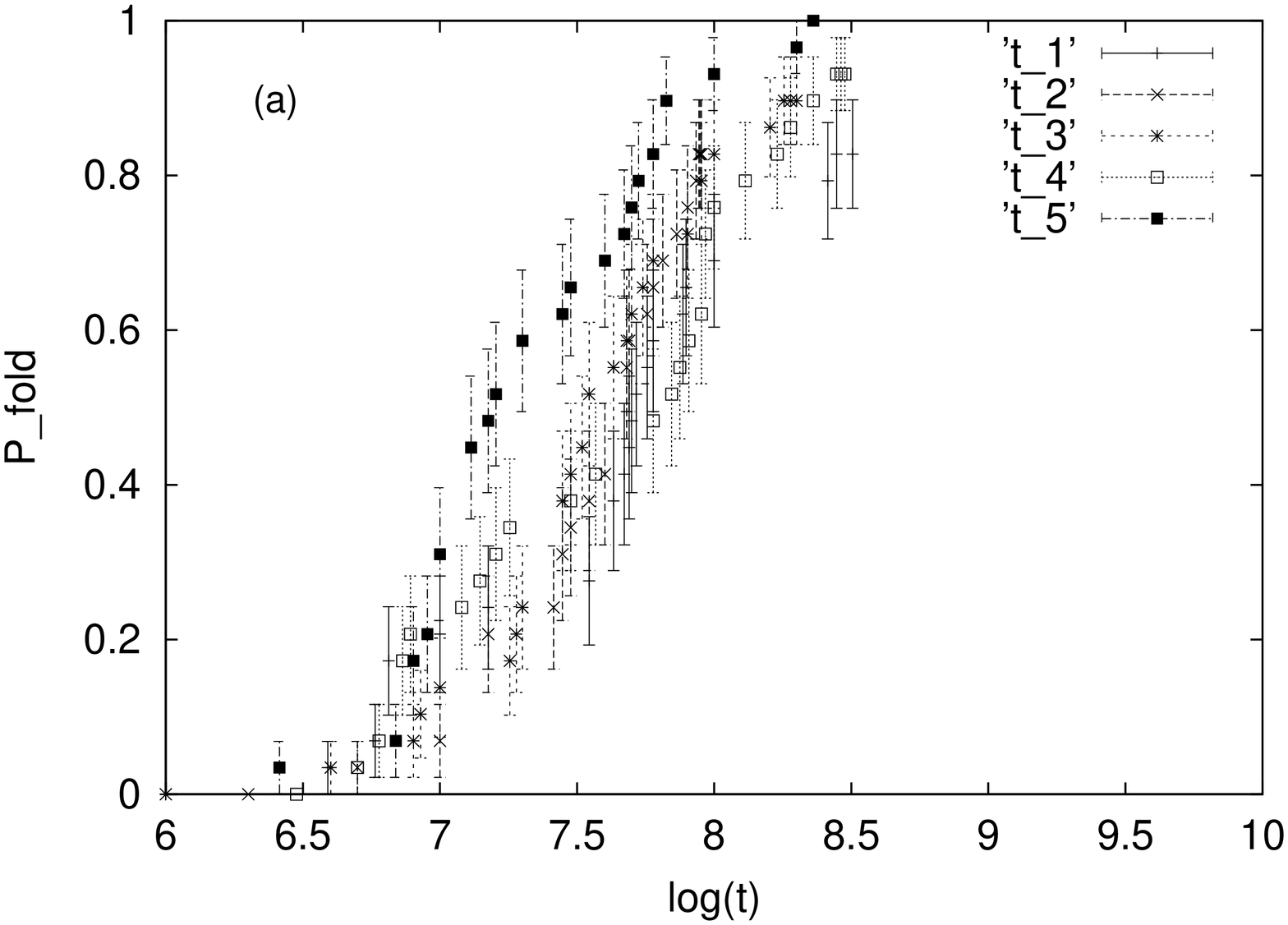, height=8cm, width=8cm, angle=270}
\psfig{file=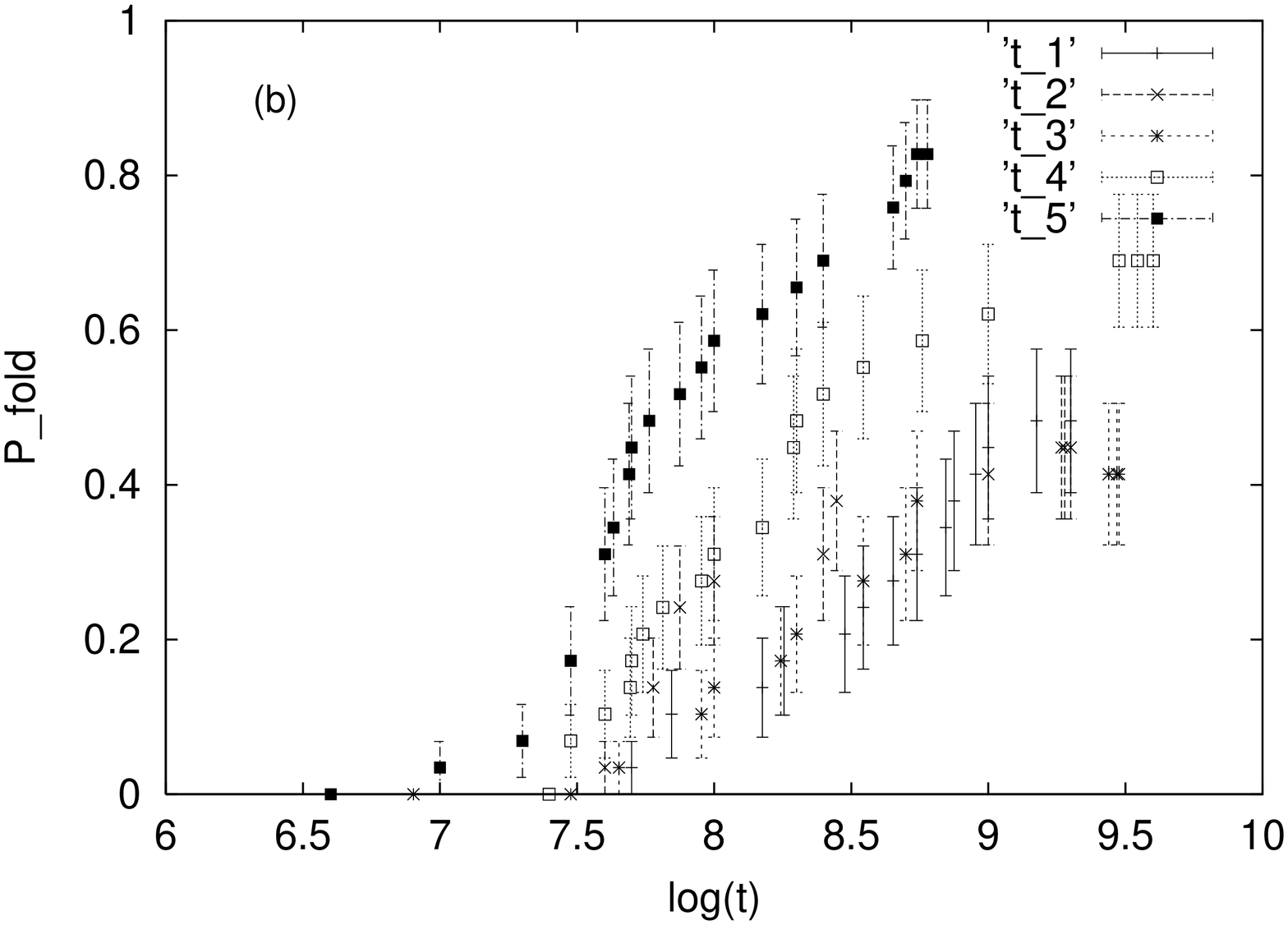, height=8cm, width=8cm, angle=270}
\caption{Separate contribution of each one of the 64 (Fig.~\ref{figure:no7}(a))
and 100 (Fig.~\ref{figure:no7}(b)) bead long targets for the
dependence of the folding probability $(P_{fold})$ on $\log(t)$ .
$P_{fold}$ was calculated as the number of folding simulations which ended up
to time $t$ normalized to the total number of runs.} 
\label{figure:no7}
\end{figure}

We have explored the time dependence of folding for the different chain lengths
$N$ each at their respective optimal folding temperature $T_{fold}(N)$. Specifically
we report in Fig.~\ref{figure:no5} the probability $P_{fold}(t)$ of the chain having
visited its target conformation after time $t$. A first look at the graph suggests
that for $N$ up to $64$ the curves appear to be functionally similar. 
A scaling factor of the form  $t^{'}=(N^{'}/N)^{\alpha}t$ translates in the
logarithmic plot to a shift
\begin{equation}
\log{t^{'}}=\log{t} + \alpha \log{\frac{N^{'}}{N}}.
\label{eq:no4}
\end{equation}
Taking $\alpha=5$ (and $N^{'}=48$) the shifted curves 
$N=27$, $36$, $48$ and $64$ superimpose well as shown in
Fig.~\ref{figure:no6}.
For $N \geq 80$ this superposition regime breaks down and the 
asymptotic value of $P_{fold}(t)$ decreases quite considerably.
Fig. ~\ref{figure:no7} shows that the break in the folding behaviour is 
associated with the onset of target-dependence of the folding curves,
where we have computed $P_{fold}(t)$ separately for each of the five targets
at both $N=64$ and $N=100$. For $N=64$ all the targets exhibit a
similar functional dependence of $P_{fold}$ on $\log(t)$. All appear consistent with
asymptotic values of $P_{fold}\rightarrow 1$ and also the dispersion of the folding time is small. However, for $N=100$ four out of the five targets have apparent asymptotics $P_{fold}\leq 1$ and
there is considerably larger dispersion of the folding time. \par
\begin{figure}
\psfig{file=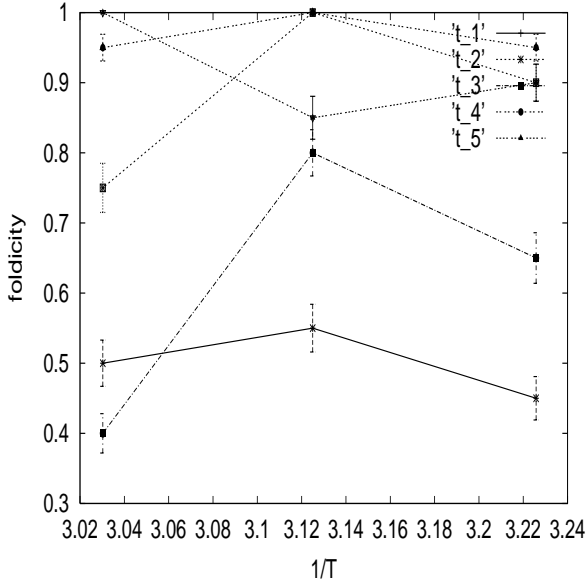, height=8cm, width=8cm, angle=270}
\caption{Dependence of foldicity on the inverse temperature, $1/T$ 
for each of the 100 bead long targets. Except for target 5, all the other targets
exhibit a maximum value of foldicity at what we considered to be the optimal folding
temperature.} 
\label{figure:no8}
\end{figure}

One possible explanation 
for this change in behaviour which we have ruled out, is that for $N \geq 80$ the optimal folding
temperature might become a target sensitive parameter. To test this hypothesis, we randomly 
selected five 100 bead long sequences (one per each target), and ran 20 folding simulations
per each value of the temperature in a certain temperature range. Fig.~\ref{figure:no8} shows how foldicity, defined as the fraction of successful folding runs over the total number of attempted runs, changes with inverse temperature for each target. Except for target 5, all the others exhibit a maximum value of foldicity at what we considered the optimal folding temperature. Target 5 marginally shows a shallow minimum and is in any case the fastest and most successful to fold in our previous results (Fig.~\ref{figure:no7}(b)),
so we can safely rule out target sensitive optimal folding temperature as the cause of target
sensitive folding performance. \par
A primary conclusion that can be drawn from these results is
that for $N \geq 80$ folding dynamics becomes target selective, certain targets
being more kinetically accessible than others. 
Having analised only five targets per chain length, we are not in a position to characterise
quantitatively the resulting distribution of behaviour. Nevertheless, the emergence of this 
new dynamical feature is a clear indication that for $N \geq 80$ thermodynamic stability is
not the dominant drive in folding, and in particular it does not solve the kinetic accessibility
problem as it has been previously suggested.

\subsection{Folding kinetics-dependence of the folding time ($t$) on the
chain length ($N$)}

\begin{figure}
\psfig{file=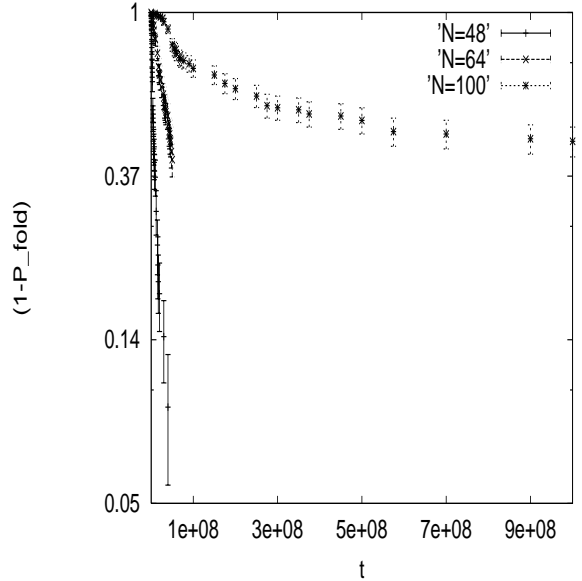, height=8cm, width=8cm, angle=270}
\caption{Evidence for a single exponential kinetic regime for
 $N=48$ and $N=64$ but not 100.}
\label{figure:no9}
\end{figure}

The folding time $t$ is a kinetic property that measures how fast a
protein sequence folds into its native state from an initial unfolded coil.
It is known that the fastest simple, single domain protein folds a million time
faster than the slowest {\cite {PLAXCO}}. However, and despite this broad
kinetic spectrum, there seems to be a general consensus among biochemists that
protein kinetics falls into two main general classes. The analysis
of experimental data collected in the course of the last ten years
has put forward the theory that proteins smaller than 100 amino acids
are committed to follow a two-state (or single exponential)
kinetics. The transition state is the only kinetically important
intermediate and conformational searching is the only factor limiting
folding speed. Bigger proteins, on the other hand
generally fold in agreement with a
mutiexponential kinetics. The latter often involves fast collapse into
kinetic traps and subsequent slower barrier climbing
out of the traps. This generates an overall process characterised by
the existence and accumulation of more than one important kinetic intermediate. 
Our results are in a broad agreement with this scenario, and the plot of $(1-P_{fold})$
in Fig.~\ref{figure:no9} supports identifying our regime for $N < 80$ with single
exponential kinetics.\par
\begin{figure}
\psfig{file=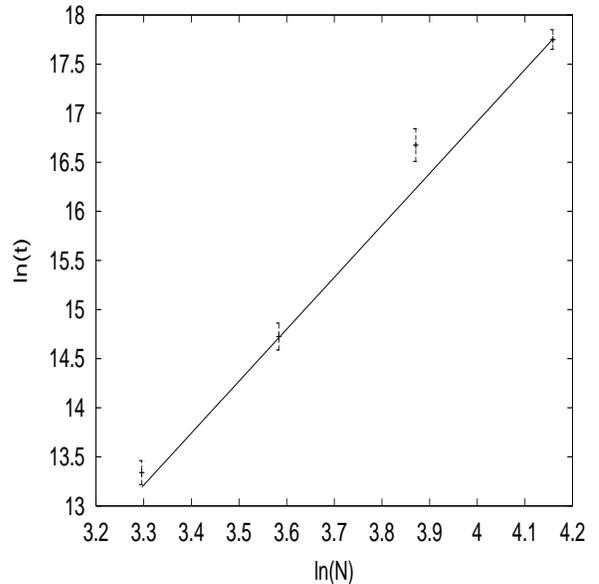, height=8cm, width=8cm, angle=270}
\caption{Dependence of the folding time $t$ on the chain length $N$ for proteins
that follow a two-state kinetics. For $N=27$, $36$, $48$ and $64$ we computed the
folding time as the mean FPT averaged over 150 simulation runs distributed over the
five considered targets. We stress that the use of different targets to compute
this plot does not smear the result because there is not target selectivity in this
folding regime.} 
\label{figure:no10}
\end{figure}

 
Recall from Fig.~\ref{figure:no6}, that tuning ${\alpha}$ to 5 in Eq.~(\ref{eq:no4})
nicely superimposes the curves of $P_{fold}$ {\it vs} $\log t$ for $N$ up to 64.
This suggests that in our first regime a scaling law
of the type $t\approx N^{5}$ appropriately describes the dependence of the
folding time, $t$, with the chain length, $N$. The plot of  
$\ln t$ {\it vs} $\ln N$ in Fig.~\ref{figure:no10} confirms (with a significant
correlation of 0.99) an exponent of 5.27. \par
Note that for $N \geq 80$ the mean FPT does not yield a correct estimate of the
folding time because the value of $P_{fold}$ does not tend to one in the limit of 
large $t$. Therefore we can only analyse the dependence of $t$ on $N$ for $N \leq 80$.\par
A scaling law of the form $t \approx N^{4}$ was suggested by a previous estimate by
Gutin {\it et al} {\cite {S1}} from folding simulations to targets which are not 
maximally compact structures. The fact that these targets have a higher kinetic
accessibility than the ones we considered might explain the weaker dependence obtained 
for the folding time on the chain length.

\subsection{Effects of stability on folding dynamics}
\begin{figure}[!h]
\psfig{file=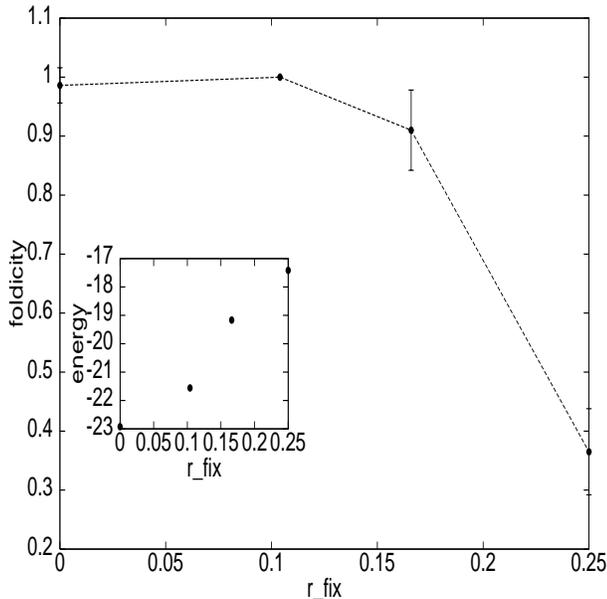, height=8cm, width=8cm, angle=270}
\caption{Dependence of foldicity on the fraction of fixed monomers,
 $r_{fix}$ (main plot) for $N=48$. The inner plot shows the dependence of the mean averaged
 sequence energy with $r_{fix}$.} 
\label{figure:no11}
\end{figure}

In this section we explore the extent to which stability, as measured by the 
sequence energy in the target affects the folding performance of proteins in our 
first regime. For this purpose we studied the MFPT to five target states of several
48 bead long sequences whose training was handicapped by fixing a priori a certain fraction
$r_{fix}$ of the beads. We designed three sets of 30 sequences per target, each set
corresponding to an $r_{fix}$ of $0.01$, $0.17$ and $0.25$ respectively.
This biases the design procedure to sequences  higher in energy as $r_{fix}$
increases allowing us to explore an energy range of $\Delta E\approx 5 $.
The MC folding simulations were again performed at $T_{fold}$ and proceeded up
to $n_{s}=9 \times 10^{8}$ MC steps or until folding was observed.
Firstly we analyse the effects of stability on foldicity.
In Fig.~\ref{figure:no11}, the main plot shows how foldicity changes with
$r_{fix}$, while in the inner plot, the averaged mean sequence energy is
plotted against the same parameter. It can be seen that foldicity is insensitive
to raising the energy up to an average value of 
$E\approx -19$ (corresponding to $r_{fix}=0.17$) but above this threshold 
it sharply decreases. We should stress however that
$E\approx -19$ is still well below the threshold heteropolymer energy, 
$E_{C}$ {\cite {S}}, for this specific chemical composition
$\langle E_{C}\rangle=-7.3849952 \pm 1.7438$ where the average is taken over
the five considered targets.\par
\begin{figure}
\psfig{file=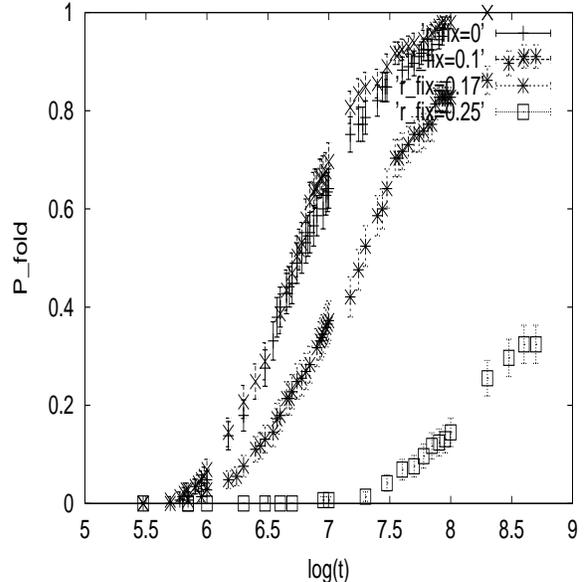, height=8cm, width=8cm, angle=270}
\caption{Dependence of the folding probability, $P_{fold}$, on $\log(t)$
for several values of the fraction of fixed monomers, $r_{fix}$.} 
\label{figure:no12}
\end{figure}
Fig.~\ref{figure:no12} shows the folding probability {\it vs} time
for each value of $r_{fix}$. 
It can be seen that while the $r_{fix}=0.17$ curve is
only shifted from the $r_{fix}=0$ and $r_{fix}=0.1$ curves (which translates in
a slower folding dynamics),
for $r_{fix}>0.17$ the folding regime clearly breaks away. Curiously, the
curves corresponding to $r_{fix}=0$
and $r_{fix}= 0.10$ nicely superpose. In energetic terms this translates
into a break in the folding regime for energies higher than $E\approx -19$.\par 
Taken together these results suggest that some degree of stability is a sufficient condition
for folding, controlling and efficiently driving the dynamics of small protein
sequences. This in turn agrees with the scenario of folding being essentially a
downhill process to the native state (energy minimum). However, designing
sequences that provide the lowest energy in the target seems to be a superflous
constraint.

\subsection{Stability and kinetic accessibility}
The picture we can draw from the results presented so far
is that for sequences whose length does not exceed 80 amino acids,
thermodinamically oriented design ensures successful folding to the targets.
As previously stated, it has been claimed {\cite {S1}} that not only do these
sequences fold stably (in the sense of allowing a high target average time
occupancy), but they also fold quicker, which suggests a correlation  
(at least to some extent) between stability and kinetic accessibility.
\par
\begin{figure}
\psfig{file=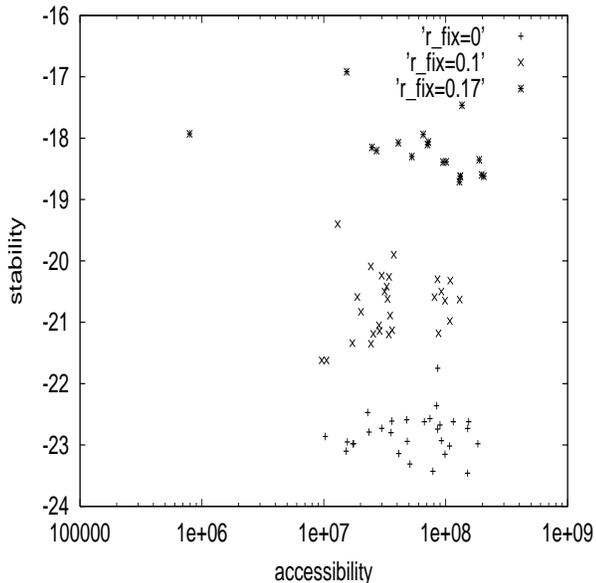, height=8cm, width=8cm, angle=270}
\caption{Accessibility-stability plot for the 48 bead long sequences whose
design stage was handicapped by fixing a priori a certain number of beads.} 
\label{figure:no13}
\end{figure}

Fig.~\ref{figure:no13} shows an accessibility-stability plot; 
accessibility is measured by the folding time $t$ (averaged mean FPT over
10 simulation runs) and
each point represents a 48 bead long sequence with energy $E$
that folds to the target in time $t$. 
\begin{figure}
\psfig{file=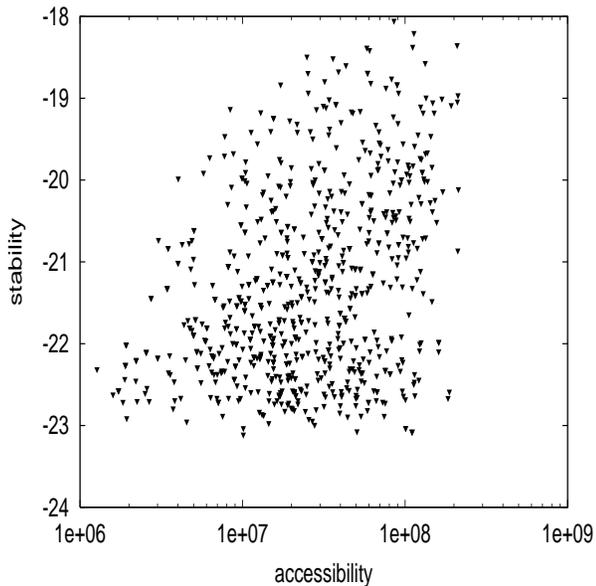, height=8cm, width=8cm, angle=270}
\caption{Accessibility-stability scatter plot for an ensemble of $\approx 1000$ 48 bead long sequences.} 
\label{figure:no14}
\end{figure}
Sequences constrained by different values of $r_{fix}$ are distinguished.

The graph strongly indicates that the quickest folders are not necessarily the most
stable sequences. In a previous report, Fink and Ball {\cite {FB}},{\cite
{FB2}}, {\cite {FB3}}.
showed through the study of a convenient ensemble of 27 bead
long sequences, that in the region of sufficient thermodynamic stability,
the latter is in conflict with optimal accessibility, 
and that a significant increase in kinetic performance
will be achieved if a marginal increase in the target's energy is allowed.\par  
In order to investigate how stability and accessibility correlate
for 48 bead long protein sequences, we prepared an ensemble of $\approx 1000$
sequences, but instead of freezing all of them,
some were annealed to some temperature different from zero. 
This allows us to scan a representative fraction of the accessibility-stability
phase space. Results are plotted Fig.~\ref{figure:no14}, where once again we
are taking $t$ as the averaged FPT over 10 simulation runs. 
The graph shows that the connection between stability and
accessibility, is not that of a simple correlation. In particular, it can be seen that
although the quickest folders appear in an energy range of high
stability ($-22< E <-23$), the most stable sequences do not show
highest accessibility.

\section{Conclusions \label{sec:4}}
The assumption that `target structure` could be an important parameter
in the dynamics of protein folding, and its subsequent introduction in the
folding simulations of a simple lattice model, made it possible to discriminate
between two distinct regimes in the dynamics. The first regime
well describes the folding of small protein molecules, and its kinetics
appears to be single exponential. In this case, conformational
search must be the only factor limiting folding speed. Folding time scales
with chain length as $t\approx N^{5}$ in this regime. On the other hand,
the folding of protein sequences bigger than 80 amino acids appears to be target
sensitive with prone to a dynamics which we might interpret as 
the falling in kinetic traps strongly delaying folding to the target. \par
The extent to which stability, as measured by the sequence energy in the target,
controls folding of proteins that fall in the first regime was investigated.
Results agree with the idea that for small protein molecules, stability is a
necessary and sufficient condition for successful folding. However,
desigining sequences for minimal energy in the target conformation
appears to be superfluous.\par
The controversial claim that the most stable sequences are also the quickest
folders was investigated.
Notwithstanding the fact that this is a delicate issue given the considerable
dimension of sequence space and the difficult task of suitably sampling it, our
results (taken from 1000 sequences) strongly indicate that the
correlation between stability and accessibility is essentially
small.\par
As a general conclusion we can say that there is much more than thermodynamics
in protein folding. In particular for long protein chains it is
evident that target geometry matters and thermodynamic factors are not the sole 
determinant of folding performance.\par
There is a growing idea among biologists, that
structural factors are a key point in protein folding dynamics.
In this context and in the scope of lattice models, it is urgent to test the 
correlation that biochemists find between CO (contact order--the average
sequence separation among contacting residue pairs) and folding rates.

\acknowledgments{The authors would like to thank Dr. Thomas M. Fink for helpful
suggestions. P.F.N. Faisca
would like to thank Programa Praxis XXI for finantial support. }

\end{document}